\newcommand{\sun}{\ensuremath{\odot}}

\documentclass[cup6a]{silverman}
\usepackage{graphicx}

\title{Star formation in galaxies hosting Active Galactic Nuclei up to $z\sim1$}

\author{John D. Silverman, zCOSMOS \& XMM-COSMOS\\ETH-Zurich, Institute of Astronomy, HIT J13.2, Wolfgang-Pauli-Strasse 27, 8093, Zurich, Switzerland}

\begin{document}
\pagenumbering{arabic}
\cleardoublepage

\maketitle
\chapter{}

\vskip -5cm 
 
\section{Introduction}

\begin{figure*}
\centering
\vspace{4cm}
\includegraphics[angle=0,scale=0.65]{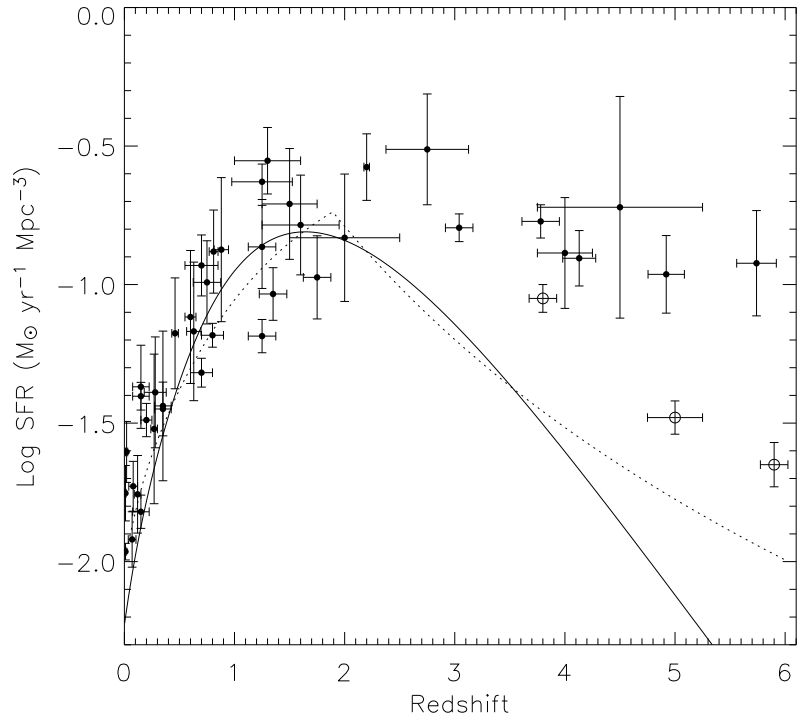}

\caption{Global evolution of star formation rate density (data points)
and mass accretion rate density onto SMBHs scaled up by a factor of
5000 for comparison (curves; See \cite{jds08a} for details).}

\label{fig1}
\end{figure*}

This contribution aims to address the fundamental question,
effectively highlighting the overall theme of the workshop, as to what
processes are important for eventually suppressing the growth of
supermassive black holes (SMBHs) and how is this related to the
evolution of star formation from $z\sim1$ to the present.  As
illustrated in Figure~\ref{fig1}, a global decline in mass accretion
onto SMBHs and star formation rate density over the last 8 Gyrs
\cite{bo98,me04,jds08a} is evident and may be driven by a mechanism
such as feedback from AGN affecting the gas content of their hosts
\cite{gr04,hop08,si09}.  Such coupling may not only explain the local
SMBH-bulge relations (see \cite{sh09} for an overview) but rectify the
inconsistency between the observed distribution of high-mass galaxies
and that predicted by semi-analytic models \cite{cr06}.

Intriguingly, there is observational evidence for AGNs influencing
their larger-scale environments that may lend support for the
aforementioned feedback models.  For example, radio jets are capable
of impacting their intracluster medium \cite{fa06} that may then
inturn regulate further cluster cooling and inhibit star formation in
the AGN host galaxy itself \cite{ra08}.  Even at low power,
radio-emitting outflows are capable of redistributing line-emitting
gas in galactic nuclei \cite{wh04}.  Furthermore, QSO-driven winds are
a common phenomenon having kinetic energies quite capable of expelling
gas especially from the nuclear region.  Although, it has not been
demonstrated that AGNs or the more luminous QSOs are responsible for quenching star formation on galactic scales.


\begin{figure*}
\centering
\vspace{1cm}
\hspace{0cm}
\includegraphics[angle=90,scale=0.5]{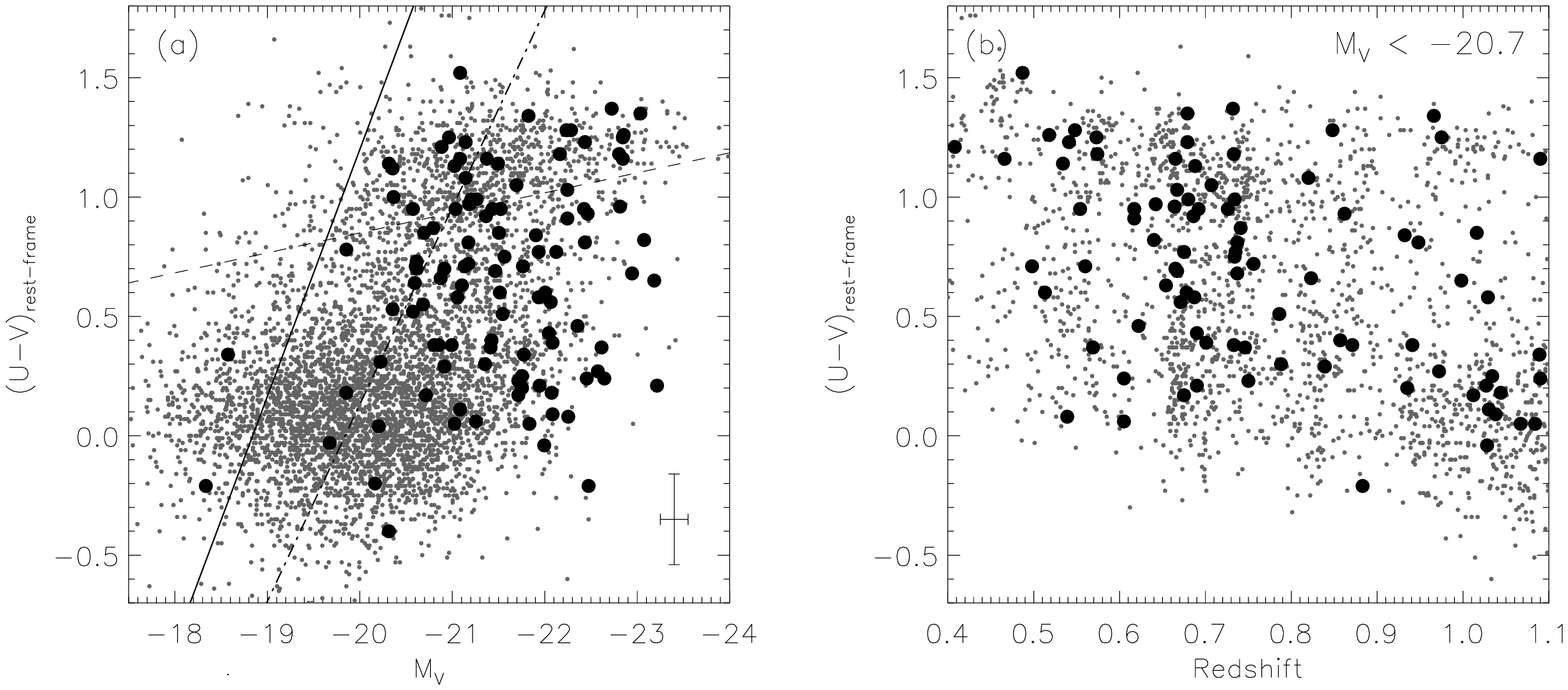}

\caption{Host galaxy colors of AGNs in the Extended $Chandra$ Deep
Field - South survey: $left$ Rest-frame color versus absolute
magnitude, $right$ Rest-frame color versus redshift.  Galaxies with
photometric redshifts from COMBO-17 \cite{wo04} are shown in grey while
those hosting X-ray selected AGN are marked by the larger black
circles.  See \cite{jds08b} for further details.}

\label{fig2}
\end{figure*}

Recent studies are providing evidence that may indicate an
impact of AGNs upon their host galaxies.  For example, X-ray selected
surveys \cite{na07,jds08b,hi09,sch09} that utilize the obscured AGN
population to enable a fairly clean view of the host galaxy
demonstrate that many have rest-frame optical colors placing them in a
region of the color-magnitude diagram usually populated with
transitional galaxies (i.e., 'green valley'; $U-V\sim0.7$; See
Figure~\ref{fig2} left panel).  This suggests that AGN feedback may
contribute to the migration of galaxies from the 'blue cloud' to the
'red sequence'.  We note that the subject of the location of AGN hosts
on the color-magnitude diagram may present an incomplete picture that
will be addressed below \cite{jds09a}.

To further explore the role of SMBHs in galaxy evolution, we aim to
determine whether AGNs are directly regulating the current rate of star
formation that may then depend on the accretion rate of the black hole
itself.  To do so, it is imperative to construct large samples of
galaxies ($>$ 10k) and compare ongoing star formation rates of
galaxies hosting AGN to those of the underlying galaxy population.
The short duty cycles of accretion onto SMBHs \cite{ma04} dictate the
need for samples of such size.  A spectroscopic nature for the sample
will further allow the disentanglement of environmental factors that
are known to influence star formation (see \cite{jds09b}).  In addition, a
multi-wavelength approach is a necessity in order to characterize the
intrinsic properties of the galaxy population (e.g., stellar mass).

The advent of such surveys starting with the Sloan Digital Sky Survey
(SDSS) are enabling investigations of the relationship between galaxy
and SMBH growth.  Based on the enormity of the SDSS database, it has
been clearly shown that galaxies hosting obscured AGN have young
stellar populations, equivalent to late-type galaxies \cite{ka03,ka07}
thus clearly establishing a direct relationship between SMBH accretion
and star formation.  On the contrary, differences between the stellar
populations of AGN hosts and galaxies lacking AGN signatures in the
SDSS have been reported \cite{ma07}, similar to the results based on
X-ray selected AGNs, and are attributed to the suppression of star
formation but may be due to selection (e.g., luminosity versus stellar mass).

Recently, large-scale spectroscopic redshift surveys (e.g., DEEP2,
COSMOS) have targeted the galaxy population out to higher redshifts
($z\sim1$) thus beginning to probe the peak of star formation and
AGN activity.  A full multi-wavelength (UV-to-IR) approach is realized
to effectively characterize the galaxy population and its evolution.
Equally important, the identification of galaxies at $z>0.3$ that host
obscured AGN demands an alternative selection technique (e.g., X-ray,
IR) to optical emission-line diagnostics.  Much progress has been made
in recent years based on these deep multiwavelength surveys to
determine the host galaxy properties of AGN (e.g.,
\cite{na07,ga09}).  In this review, I highlight our work using the
COSMOS survey with specific attention on the zCOSMOS spectroscopic
redshift survey and cospatial X-ray observations using $XMM$-Newton to
identify the galaxies, based on their stellar mass and star formation
rates, most likely to harbor an actively, accreting SMBH.  We refer
the reader to the full publication \cite{jds09a} that provide details
on the methods and analysis techniques.

\section{Star formation rates in zCOSMOS galaxies hosting AGN}

We use the zCOSMOS 10k spectroscopic redshift catalog
\cite{lilly07,lilly09} to investigate the properties of galaxies
hosting AGN and their relation to the parent population.  $XMM$-Newton
observations \cite{cap09,br07} of the full zCOSMOS sample enable us to
identify 152 AGNs that include those with significant obscuration and
of low optical luminosity.  The derived properties such as host galaxy
stellar mass, rest-frame color, and emission-line strength allow us to
determine the prevalence of AGN activity as a function of these
quantities.  Specifically, we measure the SFR of galaxies using the
[OII]$\lambda$3727 line luminosity \cite{ho05}.  We account for the contribution
from the underlying AGN component most likely arising from the
narrow-line region by using the observed [OIII]$\lambda$5007
luminosity and the typical [OII]/[OIII] ratio found from previous
studies of type 1 AGN \cite{kim06}.  The [OIII] line luminosity
is measured directly from our spectra if present.  For the subsample
with [OIII] outside our observed spectral bandpass, we infer the
[OIII] strength from the hard (2--10 keV) X-ray luminosity and the
well known correlation between these two quantities.  

\begin{figure*}
\centering
\includegraphics[angle=0,scale=0.8]{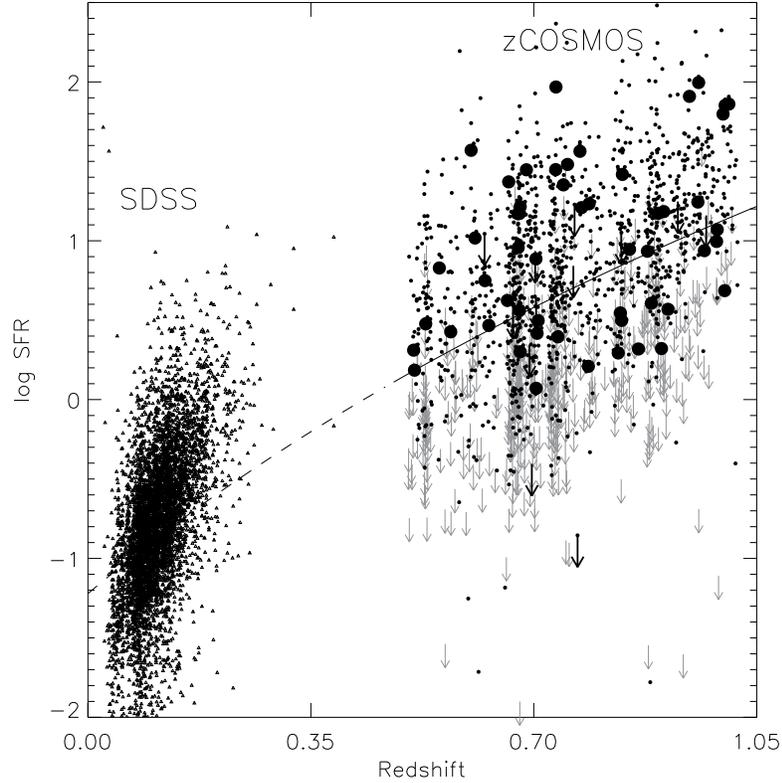}

\caption{Star formation rate versus redshift for zCOSMOS galaxies
($0.5< z < 1.0$) with those hosting AGN marked by the large black
circles.  Upper limits are shown by the arrows.  For comparison, type
2 AGNs from the SDSS are shown at $z < 0.35$ (small dots; \cite{ka03}).}

\label{fig3}
\end{figure*}

We find based on a stellar mass-selected sample of galaxies
($M_*>4\times10^{10}$ M$_\sun$) that significant levels of star
formation are present in the hosts of X-ray selected AGN
(Figure~\ref{fig3}).  SFRs (1) range from $\sim1-100$ M$_\sun$
yr$^{-1}$, with an average SFR higher than that of galaxies with
equivalent stellar mass, and (2) evolve with cosmic time in a manner
that closely mirrors the overall galaxy population.  Such evolution
appears to be consistent with the low SFRs in AGNs ($z<0.35$) from the
SDSS.  Therefore, we find no evidence for significantly reduced levels
of star formation in the hosts of AGNs and conclude that massive
galaxies with plentiful gas supplies are most conducive for AGN
activity.  This analysis effectively extends the clear association of
AGN activity and star formation, seen in low-redshift studies with the
SDSS \cite{ka03}, up to $z\sim1$.  We note that similar findings are
obtained here with additional spectral indicators, in particular
$D_n(4000)$ and rest-frame optical colors $U-V$ (see the following
section).  Finally, we highlight that these results are consistent
with the color evolution of AGN hosts, seen in the $Chandra$ Deep
Field - South survey \cite{jds08b}, that follow that of the
underlying galaxy population (See Figure~\ref{fig2} $right$ panel).

\section{Further remarks on color-magnitude diagrams of AGN hosts}

Much emphasis has been recently placed on the fact that AGN host
galaxies have rest-frame optical colors between that of blue,
star-forming galaxies and those of redder evolved galaxies (e.g.,
\cite{na07,jds08b,sch09}).  Such observations have been thought to
lend support for the role of AGNs in quenching star formation.
Although, a deficiency of AGNs within the blue galaxy population
appears to be in disagreement with the SFRs of AGN hosts in zCOSMOS
galaxies presented above and also with the fact that local ULIRGS have
not only high SFRs but enhanced levels of AGN activity \cite{sa96}.

\begin{figure*}
\centering
\includegraphics[angle=0,scale=0.7]{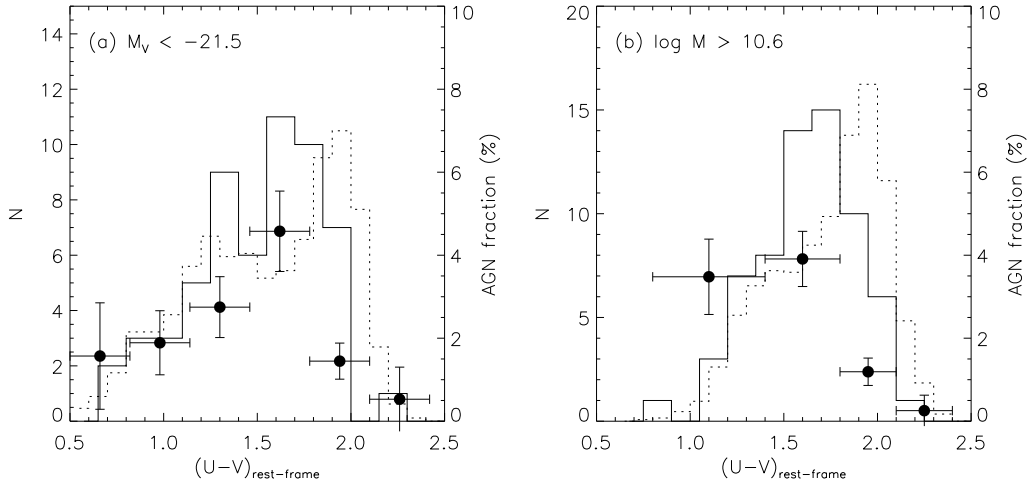}

\caption{Rest-frame color distribution of zCOSMOS galaxies (dashed histogram) and those
hosting AGN (solid histogram) for luminosity (left panel) and mass (right panel)
selected samples.  Data points show the fraction of galaxies hosting
AGN with $L_{0.5-8 {\rm keV}}\sim10^{43}$ erg s$^{-1}$.}

\label{fig4}
\end{figure*}

With this in mind, we venture further to understand why the rest-frame
colors of the hosts of AGN exhibit a difference in stellar age of
about a gigayear from that of star-forming galaxies.  It is suspected
that the discrepancy arises due to mass selection since the hosts of
type 2 AGNs in SDSS \cite{ka03} do not exhibit such a difference with
late-type galaxies of equivalent stellar mass.  To check this, we
simply determined the fraction of galaxies hosting X-ray selected AGN
as a function of rest-frame color for both a luminosity and mass
selected sample.  As shown Figure~\ref{fig4}, the difference between
the two selection methods is in the fraction of blue ($U-V<1.5$)
galaxies hosting AGN.  The mass selection mitigates the inclusion of
galaxies having low mass-to-light ratios that mainly pertains to those
forming stars.  Since the incidence of AGN activity is known to rise
with the stellar mass of its host galaxy (See Figure 7 of
\cite{jds09a}), the decline in AGN fraction from the 'green valley' to
the 'blue cloud' seen in luminosity limited samples is driven by the
preponderance of low mass galaxies that are not likely to harbor AGN
of these X-ray luminosities.  {\bf Simply put, the dependence of AGN
activity on stellar mass must be taken into consideration before
making claims regarding the color dependency of AGN activity in
luminosity-selected samples.}  In light of this check for consistency,
we conclusively find that not only our SFRs based on [OII] but the
rest-frame colors $U-V$ and spectral index $D_n(4000)$ all indicate
that AGN prefer to reside in galaxies with substantial levels of star
formation in agreement with related studies at similar redshifts
\cite{le08,al08,kiu09}.

\section{Conclusion: co-evolution of SMBHs and their host galaxies}

\begin{figure*}

\centering

\includegraphics[angle=0,scale=0.8]{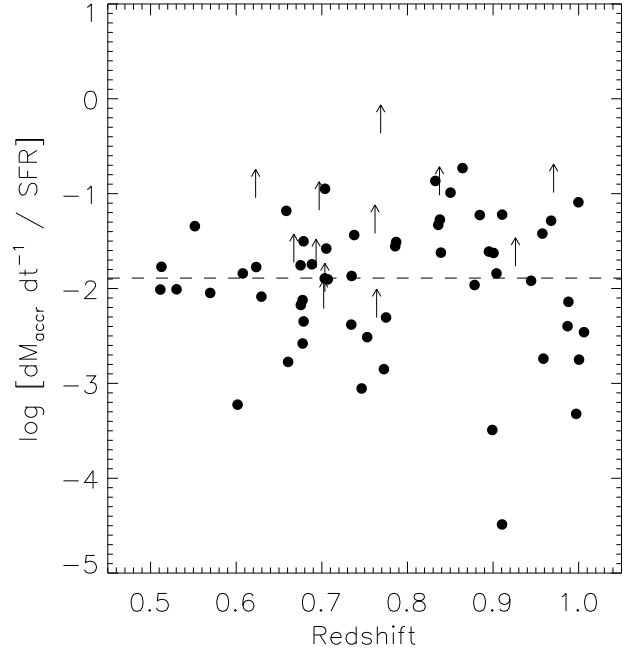}

\caption{Ratio of SMBH accretion to SFR versus redshift.  The
horizontal dashed line marks the median ratio.  Measurements are shown
by a solid circle while lower limits lower are given by an arrow.}

\label{fig5}
\end{figure*}

We have demonstrated that star formation with rates between
$\sim1-100$ M$_\sun$ yr$^{-1}$ is present in the hosts of AGN up to
$z\sim1$.  The question now is how closely does star formation track
the mass accretion rate onto these SMBHs?  To answer this question, we
have converted the X-ray luminosity to a mass accretion rate assuming
a bolometric correction and accretion efficiency.  In
Figure~\ref{fig5}, we plot the relative growth rate of these SMBHs to
their host galaxies (dM$_{accr}$ dt$^{-1}$ / SFR).  We find that a
significant amount of dispersion is present thus indicating that a
direct relationship between star formation and black hole accretion
does not occur on a case-by-case basis (See Figure 13a of
\cite{jds09a}).  On average, a co-evolution scenario is clearly evident
given the constancy of this ratio ($\sim10^{-2}$) with redshift.
Remarkably, this ratio is in very good agreement with that of low
redshift type 2 AGNs in SDSS \cite{ne09}.  The order-of-magnitude
increase in this ratio compared to the locally measured value of
$M_{BH}/M_{bulge}$, is consistent with an AGN lifetime substantially
shorter than that of star formation.  This mutual decline in global
star formation and accretion onto SMBHs, as introduced in Figure
~\ref{fig1}, is now evident within galaxies hosting AGN themselves
effectively shifting such a co-evolution scenario to smaller physical
scales.

Overall, we conclude that the properties of these X-ray selected AGN
and their host galaxies are not in accord with merger-driven models \cite{hop08} of
SMBH accretion with feedback.  Even though their SFRs are quite high,
their structural properties are not indicative of being predominantly
associated with such disturbed systems (e.g., \cite{ga09}).  Although, their hosts are massive and bulge-dominated thus
suggesting that a merger event must have happened although prior ($>$ 1 Gyr) to the AGN
phase.  As presented above, the impact of AGNs on their hosts may be minimal,
based on the levels of star formation, thus bringing into question the
efficiency of AGN feedback implemented in current semi-analytic
models.  Given the moderate-luminosities of these X-ray selected AGNs
($L_X\sim10^{43}$ erg s$^{-1}$), a "Seyfert mode" of accretion driven
by secular processes \cite{hop06} is more likely for this class of
accreting SMBHs while the more luminous QSOs \cite{ca01,ja04} may
provide the missing link to a merger-induced accretion mode.


\end{document}